\documentclass[aps,prb,twocolumn,showpacs,superscriptaddress,
preprintnumbers,amsmath,amssymb]{revtex4}
\usepackage{graphicx,subfigure}
\usepackage{dcolumn}
\usepackage{bm}
\usepackage{color}

\begin{document}

\title{Doping a Spin-Orbit Mott Insulator: Topological Superconductivity from the Kitaev-Heisenberg Model and Possible Application to (Na$_2$/Li$_2$)IrO$_3$}
\author{Yi-Zhuang You}
\affiliation{Department of Physics, University of California, Berkeley, CA 94720, USA}
\affiliation{Institute for Advanced Study, Tsinghua University, Beijing, 100084, China}
\author{Itamar Kimchi}
\affiliation{Department of Physics, University of California, Berkeley, CA 94720, USA}
\author{Ashvin Vishwanath}
\affiliation{Department of Physics, University of California, Berkeley, CA 94720, USA}
\date{\today}
\begin{abstract}
We study the effects of doping a Mott insulator on the honeycomb lattice where spins interact via direction dependent Kitaev couplings $J_\text{K}$, and weak antiferromagnetic Heisenberg couplings $J$. This model is known to have a spin liquid ground state and may potentially be realized in correlated insulators with strong spin orbit coupling. The effect of hole doping is studied within a $t$-$J$-$J_\text{K}$ model, treated using the SU(2) slave boson formalism, which correctly captures the parent spin liquid. We find  superconductor ground states with spin triplet pairing that spontaneously break time reversal symmetry. Interestingly, the pairing is qualitatively different at low and high dopings, and undergoes a first order transition with doping. At high dopings, it is smoothly connected to a paired state of electrons propagating with the underlying free particle dispersion. However, at low dopings the dispersion is strongly influenced by the magnetic exchange, and is entirely different from the free particle band structure. Here the superconductivity is fully gapped and topological,  analogous to spin polarized electrons with $p_x+ip_y$ pairing. These results may be relevant to honeycomb lattice iridates such as A$_2$IrO$_3$ (A\,=\,Li or Na) on  doping.
\end{abstract}
\maketitle

\section{Introduction}

The interplay of electron correlations and strong spin-orbit coupling (SOC) is currently attracting much attention. Mott insulators with strong SOC, such as transition metal oxides (TMO) of 5$d$ elements, can display entirely different properties from those with weak SOC, such as cuprates, manganites and nickelates\cite{OrensteinMillis}. For example, the breakdown of the spin rotation symmetry allows for magnetic Hamiltonians very different from traditionally studied SU(2) symmetric models. This can introduce a new source of frustration\cite{Balents} leading to quantum spin liquid ground states. The Kitaev honeycomb lattice model, with spin dependent interactions between spin half moments, is a remarkable example that admits an exact spin liquid ground state\cite{Kitaev}. It has recently been argued to be a natural Hamiltonian for a class of strong SOC magnets, such as the layered iridates A$_2$IrO$_3$ (A\,=\,Na, Li)\cite{Na2IrO3,Li2IrO3}, where Iridium atoms form the sites of a honeycomb lattice. In the iridium oxides, when an octahedral cage of oxygen atoms surrounds an Iridium ion, a $j=1/2$ doublet  is proposed on the Ir site\cite{KimBJ}, for which a single band Hubbard model with strong spin orbit couplings can be invoked. In the Mott insulator,  Ref.\,\onlinecite{JK,CJK} proposed that the spin couplings  include both the isotropic Heisenberg term and the strongly anisotropic Kitaev coupling:
\begin{equation}
H_{\text{HK}}=\sum _{\langle ij\rangle }J \bm{S}_i\cdot\bm{S}_j -J_\text{K} S_i^{a}S_j^{a}
\label{HK}
\end{equation}
where $S_i^{a}S_j^{a}$ is Ising coupling of the spin component $a(=1,2,3)$ according to the type of $\langle ij\rangle$ bond\cite{Kitaev} (see Fig.\,\ref{fig:lattice}).

Numerical calculations\cite{CJK, Jiang} of Eq.\,(\ref{HK}) indicate the Kitaev spin liquid phase appearing at $J=0$ persists in the range $0\leq J<J_\text{K}/8$. Although both A$_2$IrO$_3$ (A = Na, Li) are found to be magnetically ordered\cite{Na2IrO3,Li2IrO3,Hill}, their transition temperatures are relatively low. Recent experimental papers reporting magnetic susceptibility\cite{Na2IrO3, Li2IrO3} have suggested that these iridates, particularly Li$_2$IrO$_3$ may be proximate to the Kitaev spin liquid phase \cite{CJK, Jiang, finiteT}. Fits by exact diagonalization of the model Eq.\,(\ref{HK})  have reached similar conclusions\cite{kimchiyou}, but indicate that farther neighbor interactions also play a role. On the other hand, Ref.\,\onlinecite{Kim,Jin} proposed a rather different magnetic Hamiltonian, arising from large trigonal distortions, and Ref.\,\onlinecite{Nagaosa} proposed a quantum spin-Hall insulator. Future experiments should pin down the  magnetic Hamiltonian in these materials. A different realization of the Hamiltonian Eq.\,(\ref{HK}) is in perovskite iridate heterostructures of SrIrO$_3$\cite{Ran}, which produces a honeycomb lattice when grown along the (111) direction.

Motivated by these potential experimental realizations, here we will study the effects of doping the Heisenberg-Kitaev model, and investigate the conducting state that arises. To describe the physics of doping, we introduce the $t$-$J$-$J_\text{K}$ model, with the hole doping of $\delta$ per site,
\begin{equation}
H=-t\sum_{\sigma\langle ij\rangle}\mathcal{P}c^\dagger_{i\sigma}c_{j\sigma}\mathcal{P} -\mu \sum_{\sigma i}c^\dagger_{i\sigma}c_{i\sigma}+H_{\text{HK}}
\end{equation}
where the projection operator $\mathcal{P}$ removes doubly occupied sites, and the chemical potential $\mu$ is adjusted such that $\langle \sum_\sigma c^\dagger_{i\sigma}c_{i\sigma}\rangle=1-\delta$. The hopping term is nearest neighbor and spin independent. The symmetry of the honeycomb lattice along with reflection in the plane forbid a spin dependence in the nearest neighbor hopping, as evidenced by microscopic considerations\cite{Nagaosa}. Farther neighbor hoppings can be spin dependent, but are expected to be smaller and omitted in this minimal model. However, the spin-orbit interactions are nevertheless retained in the $J_\text{K}$ term. Similar Hamiltonian is also studied in Ref.\,\onlinecite{BCS,Mei}.

The $t$-$J$-$J_\text{K}$ model allows us the unique theoretical opportunity of doping a magnet which is exactly soluble in the insulating limit (at the Kitaev point), and in a spin liquid phase. The exact solution singles out the correct low energy variables --- spins represented by neutral fermions (spinons), naturally motivating a slave boson formalism. Unlike in other studies of doped Mott insulators\cite{RVB,LeeWen}, here such a formalism can be a priori justified.

Our key results are as follows: (i) Doping Kitaev spin liquid leads to a spin triplet superconductor which spontaneously breaks the time reversal symmetry. (ii) A first order transition occurs within the superconducting phase on increasing doping, which separates the two regimes SC$_1$ and SC$_2$.  In contrast,  in a similar treatment of the well known square lattice $t$-$J$ model, $d$-wave superconductivity appears  across the entire doping range at low temperature, and only quantitative properties are modified with doping. (iii) In the low doping regime (SC$_1$ phase), quasiparticle dispersions are controlled by the magnetic exchange, and leads to a time-reversal-broken triplet superconductor with the same properties as a spin-polarized $p_x+ip_y$ superconductor, which is fully gapped in the bulk but have chiral edge states and isolated Majorana modes in the vortex core\cite{ReadGreen}. This peculiar superconducting state arises because of the unusual spinon dispersion of the Kitaev spin liquid. (iv) At higher doping (SC$_2$ phase), the superconductor obtained reflects the bare dispersion of electrons, and can be smoothly connected to the weak coupling limit, where magnetic interactions lead to pairing near the Fermi surface.

This paper is structured as follows. We begin by analyzing the quantum order underlying the Kitaev spin liquid, characterized by the symmetry transformations of fractionalized excitations, a description known as 
 the projective symmetry group\cite{PSG} (PSG). We find that the Kitaev quantum order locks the spin and gauge rotations together; the two holon species transform like a spin, and spontaneously break time reversal when condensed.  
 Next we map out the mean field phase diagram within the SU(2) slave boson formalism as constrained by the Kitaev PSG, exact at zero doping,  and demonstrate that the SC$_1$ and SC$_2$ phases are dominated by different physics. Controlled by the quantum order, a time-reversal-broken triplet superconductor SC$_1$ emerges from the doped Kitaev spin liquid. We close with comments on related recent work\cite{BCS,Mei} and the relevance of our result to experimental realizations.

\section{Kitaev Spin Liquid}

To explore the physics of the $t$-$J$-$J_\text{K}$ model, we start from the well-controlled undoped and $J=0$ limit, where the model reduces to the Kitaev model. Its exact solution is given by Kitaev\cite{Kitaev} and is already well-known. Here we would like to analyze the symmetry property of the model and its spin liquid ground state.

\subsection{Symmetries of the Kitaev Model}

First, we consider the space group symmetries of the model. The symmetries are most naturally expressed by embedding the honeycomb within a 3D cubic lattice, exactly in the same manner that the Kitaev honeycomb model arises in three-dimensional  layered iridates. 
 Then the symmetry transformations, which due to spin-orbit coupling act simultaneously on spin and space, are represented in the same manner on the spin space and the 3D real space. 

\begin{figure}[htbp]
\begin{center}
\subfigure[]{\includegraphics[height=0.12\textheight]{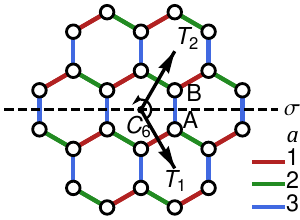}\label{fig:lattice}}\qquad
\subfigure[]{\includegraphics[height=0.12\textheight]{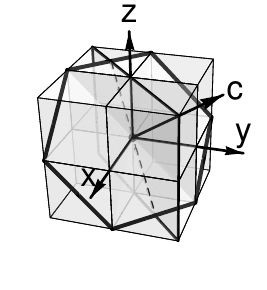}\label{fig:hexagon}}
\caption{(Color online.) (a) Lattice symmetries of the Kitaev model. The operations $C_6$ and $\sigma$ act simultaneously on lattice and spin. The three bond types ($a=1,2,3$) are colored red, green and blue respectively. (b) A hexagon plaquette embedded in the cubic lattice. The $c$-axis is the (111) axis. The 6-fold $c$-axis rotation is not a symmetry by itself, but becomes a symmetry when combined with the reflection across the lattice plane.}\label{fig:symmetry}
\end{center}
\end{figure}

Specifically, the space group is generated by two translations $T_1$ and $T_2$, an operation $C_6$ composed of a 6-fold $c$-axis rotation followed by a reflection across the lattice plane (the $c=0$ plane), and a reflection $\sigma$ across the $x=y$ plane, as illustrated in Fig.\,\ref{fig:symmetry}. 

Besides the space group symmetries illustrated above, the Kitaev model is also symmetric under time reversal $\mathcal{T}$. Time reversal has no effect on the lattice but acts as $i \sigma_2$ followed by complex conjugation $\mathcal{K}$ on the spins. While $\mathcal{T}^2=-1$ on a single spin, the global time reversal symmetry operation acting on the bipartite honeycomb lattice squares to $+1$. Combining $\mathcal{T}$ with the space group yields the full symmetry group (SG), with the presentation
$\text{SG}=\big\langle \mathcal{T},T_1,T_2,C_6,\sigma \big|
\mathcal{T}^2=1,\sigma^2=1,(C_6)^6=1\big\rangle$ 
subject to 13 definition relations, listed in Eq.\,(\ref{eq:def Cs}).

\subsection{Symmetries in a Schwinger Fermion Decomposition: the Projective Symmetry Group}
In order to study the Kitaev spin liquid and nearby phases, we must decompose the spin operator 
 $S_i^\alpha=\frac{1}{2}f_i^\dagger \sigma_\alpha f_i$ into fermionic spinons $f_i^\dagger=(f_{i\uparrow}^\dagger,f_{i\downarrow}^\dagger)$, with $\sigma_\alpha$ being the Pauli matrices. Compared to the spin operators $S_i^\alpha$, the spinon operators $f_{i\sigma}$ have an additional SU(2) gauge structure, best seen by arranging the operators into the following matrix\cite{Hermele}
\begin{equation}\label{eq:Fi}
F_i=\left(
\begin{array}{cc}
 f_{i\uparrow } & -f_{i\downarrow }^{\dagger } \\
 f_{i\downarrow } & f_{i\uparrow }^{\dagger }
\end{array}
\right).
\end{equation} 
Any right SU(2) rotation $F_i\rightarrow F_i G : G\in \text{SU(2)}$ leaves the physical spin $S_i^\alpha$ (and hence the spin Hamiltonian) unchanged, as can be seen from the following equivalent expression of $S_i^\alpha$
\begin{equation}\label{eq:S=FF}
S_i^\alpha=\frac{1}{4}\mathrm{Tr}F_i^\dagger \sigma_\alpha F_i.
\end{equation}
Therefore the right rotation $G$ corresponds to a gauge SU(2) rotation, whose generators (the SU(2) gauge charges of spinons) are given by
\begin{equation}\label{eq:K=FF}
K_i^l=\frac{1}{4}\text{Tr}\, F_i\sigma_l F_i^{\dagger }.
\end{equation}
On the other hand, the left rotation $F_i\rightarrow U^\dagger F_i: U\in\text{SU(2)}$ corresponds to the spin SU(2) rotation, whose generators are the spin operators $S_i^\alpha$.

Because of the gauge SU(2) redundancy in the Schwinger fermion representation, any SU(2) gauge operation leaves the physical spin system invariant. Any operator acting on the spins, such as a symmetry transformation, may also act within this SU(2) gauge space. Thus when we fractionalize spins in a Schwinger fermion decomposition, we must also specify how the symmetry operations of the model act within the gauge freedom. 
This extra information, known as the projective symmetry group \cite{PSG} (PSG), characterizes the fractionalized phase. 
Symmetry operations therefore consist of a symmetry group operation $g\in$SG with the corresponding spin operation $U_g$ and gauge operation $G_g$, such that the spinons transform as
\begin{equation}\label{eq:PSG}
F_i\rightarrow U_g^\dagger(i)F_{g(i)} G_g(i).
\end{equation}
The index $i$ labels the site.

In fact, the spin operation $U_g(i)=U_g$ are always site-independent, so the site index may be omitted. $U_g$'s are given by
\begin{equation}\label{eq:PSG U}
\begin{split}
&U_{T_1}=U_{T_2}=1,\\
&U_{C_6}(A)=U_{C_6}(B)=\sigma_{C_6}, \\
&U_{\sigma}(A) = U_{\sigma}(B)=\sigma _\sigma,
\end{split}
\end{equation}
where
$\sigma_{C_6} =(\sigma_0+i\sigma_1+i\sigma_2+i\sigma_3)/2$ and $\sigma_{\sigma} = i(\sigma_1-\sigma_2)/\sqrt{2}$. $\sigma_0$ is the $2\times2$ identity matrix. These matrix representations are literally translated from the descriptions of the symmetry operations on the cubic lattice, see Fig.\,\ref{fig:hexagon}. The antiunitary time reversal operation can be represented by a unitary transformation followed by a complex conjugation  $\mathcal{K}$, which transforms the spinons by
\begin{equation}\label{eq:PSG TimeRev}
F_i\rightarrow \mathcal{K}U_\mathcal{T}^\dagger(i)F_{i} G_\mathcal{T}(i) \mathcal{K},
\end{equation}
where the unitary operation acting on the spin reads
\begin{equation}\label{eq:PSG UT}
U_\mathcal{T}(A)=U_\mathcal{T}(B)=i\sigma_2.
\end{equation}
The complex conjugate operation $\mathcal{K}$ flips the sign of the imaginary unit, i.e. $\mathcal{K}i =-i\mathcal{K}$, while keeping everything else invariant ($\mathcal{K}^2=1$).

\subsection{Projective Construction for the Kitaev Spin Liquid}

The Kitaev model can be solved exactly\cite{Kitaev} by introducing 4 Majorana fermions $\chi_i^\alpha$ ($\alpha = 0,1,2,3$) on each site, and rewriting the spin operators as $S_i^\alpha = i\chi_i^0 \chi_i^\alpha$ under the constraint $\chi_i^0\chi_i^1\chi_i^2\chi_i^3=1/4$. The Majorana fermions are normalized as $\{\chi_i^\alpha,\chi_{i'}^{\alpha'}\}=\delta_{ii'}\delta_{\alpha\alpha'}$ in this work. It has been pointed out\cite{BurnellNayak} that under certain SU(2) gauge choice, the Majorana fermions $\chi_i^\alpha$ are related to the Schwinger fermions $f_{i\sigma}$ by the following matrix identity
\begin{equation}\label{eq:F=chi}
F_i = \frac{1}{\sqrt{2}}(\chi_i^0 \sigma_0+i\chi_i^1 \sigma_1+i\chi_i^2 \sigma_2+i\chi_i^3 \sigma_3),
\end{equation}
or more explicitly as $f_{i\uparrow }=\frac{1}{\sqrt{2}}(\chi _i^0+i \chi _i^3), f_{i\downarrow }=\frac{1}{\sqrt{2}}(i \chi _i^1-\chi _i^2)$. The Majorana fermions introduced by Kitaev are just another representation of the spinons. All the emergent SU(2) gauge structure for Schwinger fermions $f_{i\sigma}$ applies to the Majorana fermions $\chi_i^\alpha$ as well.

\begin{figure}[htbp]
\begin{center}
\subfigure[]{\includegraphics[width=0.16\textheight]{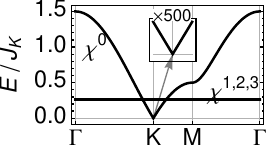}\label{fig:disp0}}
\subfigure[]{\includegraphics[width=0.16\textheight]{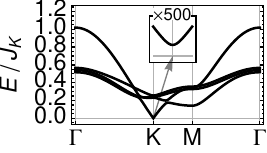}\label{fig:disp1}}
\caption{Mean field band structure of Majorana spinons (a) in the undoped limit, (b) with doping $\delta t/J_\text{K}=0.2$. The inset shows the $\times500$ zoom-in around the K point.}
\end{center}
\end{figure}

The exact ground state can be obtained by the following projective construction\cite{Wen}. First take the Majorana bilinear Hamiltonian 
\begin{equation}\label{eq:KitaevMF}
H=J_K\sum_{\langle ij\rangle}\left(iu_{ij}^a\chi_i^0\chi_j^0+iu_{ij}^0\chi_i^a\chi_j^a -u_{ij}^0 u_{ij}^a\right),
\end{equation}
where the bond parameters $u_{ij}^\alpha=\langle i\chi_i^\alpha\chi_j^\alpha\rangle$ ($\alpha=0,1,2,3$) can be regarded as the mean field ansatz, self-consistently given by
\begin{equation}\label{eq:Kitaev ansatz}
u_{ij}^\alpha = 
 \left\{ 
\begin{array}{ll}
 -0.262433 & \text{if }\alpha =0, \\
 1/2 & \text{if }\alpha =a, \\
 0 & \text{otherwise}.
\end{array}\right.
\end{equation}
Here $a$ denotes the type of the bond $\langle i j\rangle$. We choose $i\in A$ sublattice and $j\in B$ sublattice to be the positive bond direction. Given the ansatz Eq.\,(\ref{eq:Kitaev ansatz}), the mean field Hamiltonian Eq.\,(\ref{eq:KitaevMF}) produces a graphene-like band structure for $\chi^0$ and degenerate flat bands for $\chi^1$, $\chi^2$ and $\chi^3$, as shown in Fig.\,\ref{fig:disp0}. Take the Majorana Fermi liquid ground state and project to the physical Hilbert space by imposing the condition $\chi_i^0\chi_i^1\chi_i^2\chi_i^3=1/4$, the resulting state is the exact ground state given by Kitaev. 

The spin correlation in this state was shown to be short-ranged\cite{Baskaran}, which identifies the ground state of the Kitaev model as a quantum spin liquid. However what really differentiates the spin liquid from a trivial spin disordered paramagnetic state is the quantum order\cite{quantum order} encoded in the Majorana Fermi liquid from which the spin liquid is obtained by projection. Given the particular mean-field ansatz parameterized by $u_{ij}^\alpha$, the $\chi^0$ fermion has a band structure different from $\chi^{1,2,3}$, so it is no longer possible to mix $\chi^0$ with the other Majorana fermions. Thus the emergent SU(2) gauge structure of mixing spinon flavors is broken down to the $\mathbb{Z}_2$ gauge structure of changing sign of $\chi^\alpha$. The broken gauge structure can be imagined as a hidden order of spinon superconductivity\cite{BurnellNayak}. Although it will not manifest as electron superconductivity in the spin liquid due to the lack of charge fluctuation, its existence as a quantum order is real, and will be revealed, once the charge fluctuation is introduced by doping. 

\subsection{Projective Symmetry Group of the Kitaev Spin Liquid}

More precisely, the quantum order\cite{quantum order} of the $\mathbb{Z}_2$ spin liquid is characterized by the PSG of the mean field ansatz. The PSG of Kitaev spin liquid can be determined starting from the fact that $\chi^0$ is a special flavor which should not be mixed with other flavors, any PSG operation must at least preserve the flavor of $\chi^0$.  $\chi^0$ appears in the $F$ matrix as $F\sim\chi^0\sigma_0$, while $F$ transforms under PSG operations as $F\to U_g^\dagger F G_g$, so apart from some sign factor, $\chi^0\sigma_0\to\pm\chi^0 U_g^\dagger G_g$. Therefore, to preserve the flavor of $\chi^0$, $G_g=\pm U_g$ is simply required to hold for all $g\in\text{SG}$: the gauge operation $G_g$ must always follow the spin operation $U_g$ up to a sign factor. From the spin operations $U_g$ given in Eq.\,(\ref{eq:PSG U}) and  Eq.\,(\ref{eq:PSG UT})  it is not difficult to figure out the gauge operations $G_g$, which read
\begin{equation}\label{eq:PSG G}
\begin{split}
&G_{T_1}=G_{T_2}=1,\\
&G_{C_6}(A)=-G_{C_6}(B)=\sigma_{C_6}, \\
&G_{\sigma}(A) = -G_{\sigma}(B)=\sigma _\sigma,\\
&G_{\mathcal{T}}(A) = -G_{\mathcal{T}}(B)=i\sigma _2.
\end{split}
\end{equation}
The matrices $\sigma_{C_6}$ and $\sigma_{\sigma}$ were defined right below Eq.\,(\ref{eq:PSG U}). The sublattice-dependent sign factors are determined as follows. Both $C_6$ and $\sigma$ switch the sublattice $A$ and $B$, carrying $u_{AB}^\alpha$ to $u_{BA}^\alpha$ under the lattice transformation. However, $u_{ij}^\alpha=-u_{ji}^\alpha$ is odd under the reversal of bond direction, so in order to keep it unchanged, the sign must be rectified by the gauge operation that follows, therefore both $G_{C_6}$ and $G_{\sigma}$ have a sign difference between the sublattices. However for the time reversal operation, under complex conjugate $i\to-i$, so $u_{AB}^\alpha=\langle i\chi_A^\alpha\chi_B^\alpha\rangle\to\langle -i\chi_A^\alpha\chi_B^\alpha\rangle=-u_{ij}^{\alpha}$, thus the gauge transform $G_\mathcal{T}$ must also carry the sublattice-dependent sign to compensate the sign generated by the complex conjugate.

A prominent property of the PSG of the Kitaev spin liquid is that $U_g$ and $G_g$ are always the same (up to a sign), which implies that the spin and gauge degrees of freedom are locked together by the underlying quantum order in the spin liquid state. As a result, the PSG operation $U_g^\dagger F G_g$ literally carries out the rotations and reflections by treating $\chi^0$ as a scalar and $\bm{\chi}\equiv(\chi^{1},\chi^{2},\chi^{3})$ as a pseudo vector. Therefore $C_6$ actually permutes $\chi^3\rightarrow \chi^2 \rightarrow \chi^1 \rightarrow \chi^3$, and $\sigma$ exchanges $\chi^1\leftrightarrow\chi^2$, with some additional sign factors (see Tab.\,\ref{tab:PSG}), thus giving exactly the right transforms to preserve all the mean field ansatz, which can be checked straightforwardly.

\begin{table}[htdp]
\caption{The PSG transforms of Majorana fermions.}\label{tab:PSG}
\begin{center}
\begin{tabular}{rrrrrr}
$g:$&$T_{1,2}$& $C_6$ & $\sigma$& $\mathcal{T}$\\
\hline
$\chi_A^0\rightarrow$&$\chi_A^0$&$\chi_B^0$&$\chi_B^0$&$\chi_A^0$\\
$\chi_A^1\rightarrow$&$\chi_A^1$&$\chi_B^3$&$-\chi_B^2$&$\chi_A^1$\\
$\chi_A^2\rightarrow$&$\chi_A^2$&$\chi_B^1$&$-\chi_B^1$&$\chi_A^2$\\
$\chi_A^3\rightarrow$&$\chi_A^3$&$\chi_B^2$&$-\chi_B^3$&$\chi_A^3$\\
\hline
$\chi_B^0\rightarrow$&$\chi_B^0$&$-\chi_A^0$&$-\chi_A^0$&$-\chi_B^0$\\
$\chi_B^1\rightarrow$&$\chi_B^1$&$-\chi_A^3$&$\chi_A^2$&$-\chi_B^1$\\
$\chi_B^2\rightarrow$&$\chi_B^2$&$-\chi_A^1$&$\chi_A^1$&$-\chi_B^2$\\
$\chi_B^3\rightarrow$&$\chi_B^3$&$-\chi_A^2$&$\chi_A^3$&$-\chi_B^3$
\end{tabular}
\end{center}
\label{default}
\end{table}%

In conclusion, the PSG of the Kitaev spin liquid is defined by Eq.\,(\ref{eq:PSG}) in general (and by Eq.\,(\ref{eq:PSG TimeRev}) for the time reversal operation), with the spin and gauge transforms specified by Eq.\,(\ref{eq:PSG U}), Eq.\,(\ref{eq:PSG UT}) and Eq.\,(\ref{eq:PSG G}). Its effect on the Majorana spinons are concluded in Tab.\,\ref{tab:PSG}. This PSG belongs to the class (I)(B) according to the PSG classification of $\mathbb{Z}_2$ spin liquid on the honeycomb lattice (see Appedix \ref{app:PSG} for details of the classification).

All the PSG's in this class have the common property that the gauge charge is reversed under time reversal just the same as the spin. To see this, substitute Eq.\,(\ref{eq:F=chi}) into Eq.\,(\ref{eq:S=FF}) and Eq.\,(\ref{eq:K=FF}), and write the spin and gauge charge operators in terms of Majorana fermions as
\begin{equation}
\begin{split}
\bm{S}_i&=\frac{i}{2}\big(\chi _i^0\bm{\chi }_i-\frac{1}{2}\bm{\chi }_i\times \bm{\chi }_i\big),\\
\bm{K}_i&=\frac{i}{2}\big(\chi _i^0\bm{\chi }_i+\frac{1}{2}\bm{\chi }_i\times \bm{\chi }_i\big),
\end{split}
\end{equation}
Applying the PSG transformation rules of the time reversal: $\chi_A^\alpha\rightarrow\chi_A^\alpha$, $\chi_B^\alpha\rightarrow-\chi_B^\alpha$ (see Tab.\,\ref{tab:PSG}) and $i\rightarrow -i$, it is easy to show that both the spin and gauge charge operators are odd under time reversal
\begin{equation}
\bm{S}_i\overset{\mathcal{T}}{\to}-\bm{S}_i, \bm{K}_i\overset{\mathcal{T}}{\to}-\bm{K}_i.
\end{equation}
Therefore, there are in principle two ways to to break the time reversal symmetry in the Kitaev spin liquid: one is to polarize the spin and the other is to condense the gauge charge. The spin polarization can be achieved by applying an external magnetic field in the (111) direction (perpendicular to the lattice plane), which drives the gapless Kitaev spin liquid into the gapped non-Abelian phase\cite{Kitaev,Jiang}. In the following, we will explore the second possibility, namely the gauge charge condensation. This can be achieved by introducing the gauge charge through doping the spin liquid. According to the SU(2) slave boson theory, the condensed holon will pick out an SU(2) gauge direction and break the time reversal symmetry spontaneously.

\section{Doping the Kitaev model within SU(2) Slave Boson Theory}

\subsection{SU(2) Slave Boson / Schwinger Fermion Representation}
We now consider doping (say) holes into the insulating magnet, while preserving the strong onsite correlations that penalize double occupancy. As discussed above, the exact solution of the Kitaev spin liquid is naturally expressed within a particular kind of Schwinger fermion / slave boson representation. The most naive way is to directly assign the spinons to the electrons $c_{i\sigma}=b_i f_{i\sigma}$ with additional U(1) slave boson $b_i$ to carry the electric charge. However this approach completely neglects the SU(2) gauge redundancy in the spin liquid: annihilation of a spin up electron by $c_{\uparrow}$ can be accomplished (in the spin sector) either by the annihilation of up spinon $f_{\uparrow}$ or by the creation of down spinon $f_{\downarrow}^\dagger$ (to neutralize the up spin into spin singlet), so the electron operator must be a linear combination of both\cite{LeeWen, LeeRMP}, formulated as $c_{i\uparrow }=\frac{1}{\sqrt{2}}(b_{i1}^{\dagger }f_{i\uparrow }-b_{i2}^{\dagger }f_{i\downarrow }^{\dagger })$, $c_{i\downarrow }=\frac{1}{\sqrt{2}}(b_{i1}^{\dagger }f_{i\downarrow }+b_{i2}^{\dagger }f_{i\uparrow }^{\dagger })$, or equivalently as\cite{Hermele}
\begin{equation}\label{eq:C=FB}
C_i=\frac{1}{\sqrt{2}}F_iB_i,
\end{equation} 
where $C_i$, $F_i$ and $B_i$ are $2\times2$ matrices of operators
\begin{equation}
C_i=\left(
\begin{array}{cc}
 c_{i\uparrow } & -c_{i\downarrow }^{\dagger } \\
 c_{i\downarrow } & c_{i\uparrow }^{\dagger }
\end{array}
\right),
B_i=\left(
\begin{array}{cc}
 b_{i1}^{\dagger } & -b_{i2} \\
 b_{i2}^{\dagger } & b_{i1}
\end{array}
\right),
\end{equation}
and $F_i$ is given by Eq.\,(\ref{eq:Fi}) in terms of Schwinger fermions or equivalently by Eq.\,(\ref{eq:F=chi}) in terms of Majorana fermions. The holon creation operators $b_{i1}^\dagger$ and $b_{i2}^\dagger$ carry different SU(2) gauge charges, but the same electric charge as a hole $c_i$.
 
Let $|0\rangle_\text{slave}$ be the vacuum state of both spinons and holons, s.t. $f_{i\sigma}|0\rangle_\text{slave}=b_{i\nu}|0\rangle_\text{slave}=0$. Then on each site, there are only three physical states in the Hilbert space:
\begin{equation}\label{eq:physical states}
\begin{split}
|0\rangle &=\frac{1}{\sqrt{2}}\left(b_{i1}^{\dagger }+b_{i2}^{\dagger }f_{i\uparrow }^{\dagger }f_{i\downarrow }^{\dagger }\right)|0\rangle _{\text{slave}},\\
c_{i\uparrow }^{\dagger }|0\rangle &=f_{i\uparrow }^{\dagger }|0\rangle _{\text{slave}},\\
c_{i\downarrow }^{\dagger }|0\rangle &=f_{i\downarrow }^{\dagger }|0\rangle _{\text{slave}}.
\end{split}
\end{equation}
Here $|0\rangle$ denotes the electron empty state. The double occupied state is  automatically ruled out from the physical Hilbert space in the SU(2) slave boson formalism.

Each empty site has one holon, therefore the doping $\delta$ is: $\delta=\frac1N\sum_i(b_{i1}^{\dagger }b_{i1}+b_{i2}^{\dagger }b_{i2})$, where $N$ denotes the total number of sites. Adopting Gutzwiller approximation, the spin operator will be written as $S_i^a=i\chi_i^0\chi_i^a(1-\delta)$.

\subsection{SU(2) Gauge Charge}
As both spinons and holons carry the SU(2) gauge charges, the gauge SU(2) generators $K_i^l$ ($l=1,2,3$) are generalized from Eq.\,(\ref{eq:K=FF}) to
\begin{equation}
K_i^l=\frac{1}{4}\text{Tr}\, F_i\sigma_l F_i^{\dagger }-\frac{1}{4}\text{Tr}\, \sigma _3B_i^{\dagger }\sigma_l B_i,
\end{equation}
or explicitly written as (with implicit sum over dummy indices)
\begin{equation}\label{eq:K}
K_i^l=-\frac{1}{2}(i\chi_i^0\chi_i^l+\frac{i}{2}\epsilon_{lmn}\chi_i^m\chi_i^n+b_{i\nu}\sigma^l_{\nu\nu'}b_{i\nu'}^\dagger),
\end{equation}
where $\epsilon_{lmn}$ is the Levi-Civita symbol. It can be verified that $[K_i^l,F_i]=\frac{1}{2}F_i \sigma_l$, $[K_i^l,B_i]=-\frac{1}{2}\sigma_l B_i$, therefore $[K_i^l,C_i]=0$, showing that $K_i^l$ are indeed the generators of gauge SU(2) transforms that leave the electron operators unchanged.

The physical state, as enumerated in Eq.\,(\ref{eq:physical states}), are SU(2) gauge invariant. Therefore the SU(2) singlet condition $K_i^l=0$ should be imposed. This condition is equivalent to the single occupancy condition for both spinons and holons, as is evidenced from $K_i^3=(1-f_{i\uparrow}^\dagger f_{i\uparrow}-f_{i\downarrow}^\dagger f_{i\downarrow}-b_{i1}^{\dagger }b_{i1}+b_{i2}^{\dagger }b_{i2})/2=0$.

The PSG operations are naturally extended to the holons, such that they transform as
\begin{equation}
\begin{split}
B_i&\rightarrow G_g^\dagger(i)B_{g(i)},\quad\text{(for $g\neq\mathcal{T}$)}\\
B_i&\rightarrow \mathcal{K}G_\mathcal{T}^\dagger B_{i}\mathcal{K}.
\end{split}
\end{equation}
In particular, under the time reversal operation,
\begin{equation}
\left(
\begin{array}{c}
 b_{A1} \\
 b_{A2}
\end{array}
\right)
\overset{\mathcal{T}}{\to }
\left(
\begin{array}{c}
 -b_{A2}\\
 b_{A1}
\end{array}
\right).
\end{equation}
One can see the holon SU(2) gauge charges transform under time reversal in a way similar to the physical spins. Therefore one could expect that the condensation of holons will spontaneously breaks the time reversal symmetry.

\subsection{Mean Field Phase Diagram}
The exact solution of the Kitaev spin liquid at zero doping involves an enlarged hilbert space with spinons and holons which implements a particular PSG. We expect these deconfined excitations, which transform under symmetry operations as defined by the Kitaev-limit PSG, to survive into finite doping. At small finite doping the SU(2) slave boson mean field with this particular PSG becomes inexact, but should still provide the most accurate treatment possible.

Using Eq.\,(\ref{eq:C=FB}), the $t$-$J$-$J_\text{K}$ model can be written in terms of spinons and holons (for simplicity we set $J=0$, finite $J$ is discussed in Appendix \ref{app:finiteJ}). 
Then use the mean field treatment by introducing the following mean field parameters:
\begin{equation}\label{eq:uw}
u_{ij}^\alpha= \langle i \chi_i^\alpha \chi_j^\alpha \rangle,\quad
w_{ij}^\nu= \langle i b_{i\nu}^\dagger b_{j\nu} \rangle,
\end{equation}
we arrive at the mean field Hamiltonian (see Appendix \ref{app:MF} for detailed deductions)
\begin{equation}\label{eq:HMF}
\begin{split}
H_{\text{MF}}=&\sum_{\langle i j\rangle} U_{ij}^\alpha i \chi_i^\alpha \chi_j^\alpha +  W_{ij}^\nu (i b_{i\nu}^\dagger b_{j\nu}+h.c.)\\
&+\sum_i a_i^l K_i^l-\mu b_{i\nu}^\dagger b_{i\nu},
\end{split}
\end{equation}
where summation is implied over repeated indices $\alpha=0\dots 3$, $\nu=1,2$ and $l=1\dots 3$. The hopping amplitudes for fermions $U_{ij}^\alpha$ and for bosons $W_{ij}^\nu$ should be determined self-consistently from 
\begin{equation}
\begin{split}
U_{ij}^\alpha = & -\frac{t}{4}\sum_{\nu=1}^2(w_{ij}^{\nu}+c.c.) \\ &+J_K (1-\delta)^2 (u_{ij}^a\delta_{0\alpha}+u_{ij}^0\delta_{a\alpha}),\\
W_{ij}^\mu = & -\frac{t}{4}\sum_{\alpha = 0}^3 u_{ij}^\alpha.
\end{split}
\end{equation}
The index $a$ denotes the direction of $\langle ij \rangle$. The boson chemical potential $\mu$ is chosen such that $\sum_{i,\nu}\langle b_{i\nu}^{\dagger }b_{i\nu}\rangle =\delta N$.
The SU(2) gauge charge operators are given in Eq.\,(\ref{eq:K}). The gauge potentials $a_i^l$ are chosen to enforce the SU(2) gauge singlet constraint on average $\langle K_i^l\rangle = 0$. In the undoped limit, Eq.\,(\ref{eq:HMF}) reduces to the mean field description of the spin liquid exact solution. With finite doping, the hidden superconductivity of spinons will be rendered into the true superconductivity of electrons once the holons condense.

We would like to stress that the quantum order of the Kitaev spin liquid puts a strong constraint on the possible form of the mean field ansatz. This quantum order is described by the Kitaev spin liquid PSG as discussed previously. We assume that this PSG is respected by the mean field solution throughout, and that symmetry breaking occurs only through holon condensation. At small dopings this is required by continuity to the Kitaev solution. The most general parameterization of the mean field ansatz under the PSG restriction is as follows. First assign on the type-3 bond
\begin{equation}
\begin{split}
&u_{ij}^0= u_0, u_{ij}^1=u_{ij}^2=u_b, u_{ij}^3=u_a,\\
&w_{ij}^1=w_{ij}^2= w.
\end{split}
\end{equation}
Then the mean field parameters on the other bonds are obtained by using the PSG operation to carry the above assignment throughout the lattice. $u_0$, $u_a$, $u_b$ and $w$ are real numbers that parameterize the mean field ansatz. 

\begin{figure}[htbp]
\begin{center}
\includegraphics[width=0.24\textheight]{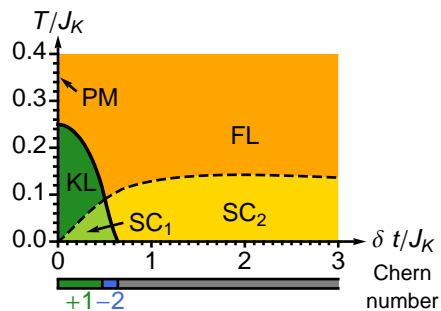}
\caption{(Color online) Mean field phase diagram for $t=10J_\text{K}$ and $J=0$. The low doping Kitaev spin liquid (KL) phase and the high doping Fermi liquid (FL) phase are separated by first order transition. Once holons condense, two classes of superconducting (SC$_1$ and SC$_2$) phases appear. The bar below shows the Chern number of the superconducting state.}
\label{fig:phase}
\end{center}
\end{figure}

Based on the parameterization, a self-consistent mean field solution of Eq.\,(\ref{eq:HMF}) gives the phase diagram shown in Fig.\,\ref{fig:phase}. We show results for $J=0$. Introducing $J<J_\text{K}/8$, to remain within the boundary of the spin liquid phase\cite{CJK, Jiang}, has little effect on the phase diagram (discussed in Appendix \ref{app:finiteJ}).

\subsection{Spin Liquid and Adjacent Phases}

In the undoped limit, one recovers the Kitaev spin liquid mean field parameters: $u_b=w=0$, and $u_0$ and $u_a$ are determined by the following self-consistent equations
\begin{equation}
\begin{split}
u_a &=-\frac{1}{2}\tanh \frac{\beta  J_K u_0 }{2},\\
u_0 &=-\frac{1}{3N}\sum _{\bm{k}\in \text{BZ}} |\Gamma (\bm{k})|\tanh \frac{\beta J_K u_a  |\Gamma (\bm{k})|}{2},
\end{split}
\end{equation}
where $\Gamma(\bm{k})=e^{ik_y}+2e^{-ik_y/2}\cos(\sqrt{3}k_x/2)$, and $N$ is the number of sites. At zero temperature, the solution is $u_0 =-0.262433$ and $u_a =1/2$, corresponding to the exact ground state of the Kitaev model. So the SU(2) slave boson mean field theory is asymptotically exact in the small doping limit. At the mean field level, a finite temperature transition is found at $T_c = J_K/4$, above which ($T>T_c$) all the mean field parameters vanish, $u_0=u_a=u_b=w=0$. The confining gauge fluctuation will recombine spinons and holons into electrons, resulting in a paramagnetic (PM) phase.

\begin{figure}[htbp]
\begin{center}
\includegraphics[height=0.11\textheight]{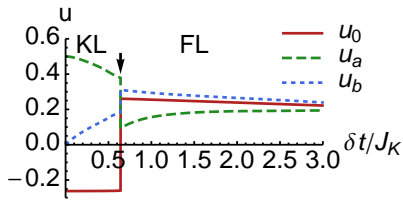}
\caption{(Color online.) Mean field parameters $u_0$, $u_a$, $u_b$ v.s. doping $\delta$ at zero temperature. The arrow indicates the 1st order transition between KL/SC$_1$ phase and FL/SC$_2$ phase. The calculation is done at $t=10J_\text{K}$ and $J=0$.}
\label{fig:upara}
\end{center}
\end{figure}

With increasing doping, mean field parameters $u_b$ and $w$ grow in proportional to $\delta$, and eventually trigger a first order phase transition at $\delta_c\simeq 2u_a J_\text{K}/t$, see Fig.\,\ref{fig:upara}. The transition is driven by the competition between the kinetic energy of holes ($t$ term) and the magnetic energy of spins ($J_\text{K}$ term). The magnetic energy favors the Kitaev spin liquid state, in which the mobility of $\chi^{1,2,3}$ fermions is sacrificed (as they form degenerate flat bands). For larger doping, more kinetic energy can be gained by allowing $\chi^{1,2,3}$ fermions to move in the same way as $\chi^0$, as $u_0 \simeq u_a\simeq u_b$, so that the flat band gets dispersed as shown in Fig.\,\ref{fig:disp1}. In the large doping limit, all flavors of Majorana fermions move with the same amplitude, providing identical graphene-like band structures, which can be recombined into band electrons, labeled by Fermi liquid (FL) in Fig.\,\ref{fig:phase}. As discussed below, the nature of superconductivity is very different depending on the normal state, Kitaev spin liquid or FL, from which it emerges.

\subsection{Holon Condensation and Superconductivity}

At low temperature, the holons condense to their band minimum at zero-momentum, leading to the following condensate amplitude ($\nu=1,2$) 
\begin{equation}\label{eq:b=z}
\langle b_{A\nu}\rangle=z_\nu, \langle b_{B\nu}\rangle=iz_\nu,
\end{equation}
with the density $|z_1|^2+|z_2|^2=\delta$ following the doping level. The electron pairing is found between opposite sublattices (because the intra-sublattice coupling of $\chi^\alpha$ is forbidden by PSG):  $\Delta_{AB,b}(\bm{k})=c_{\bm{k}A}^{\intercal}\epsilon\sigma_b c_{-\bm{k}B}$, ($b=0,1,2,3$) where $c_{\bm{k}A(B)}=(c_{\bm{k}A(B)\uparrow}, c_{\bm{k}A(B)\downarrow})^{\intercal}$ denote the electron operators in the momentum space and $\epsilon=i\sigma_2$ is the anti-symmetric matrix\cite{Volovik}. Using Eq.\,(\ref{eq:C=FB}), the pairing is expressed in terms of the mean field parameters: 
\begin{equation}
\Delta_{AB,b}(\bm{k})=\frac{\alpha_b}{2}\sum _{a=1}^3 d_{a b}e^{i \bm{k}\cdot \bm{r}_a},
\end{equation}
where  $\bm{r}_1=(-\sqrt{3}/2,-1/2)$, $\bm{r}_2=(\sqrt{3}/2,-1/2)$, $\bm{r}_3=(0,-1)$ denote the three displacement vectors from site $A$ to site $B$, and $b$ labels the singlet ($b=0$) or triplet ($b=1,2,3$) channels. $\alpha_b=z^\intercal\epsilon \sigma_b z$ with $z =(z_1, z_2)^\intercal$ refer to the holon condensate amplitude, and $d_{ab}=u_0-u_a+2(u_a-u_b)\delta_{ab}$ parameterize the the spinon pairing amplitude. The electron superconductivity is a joint effect of holon condensation and spinon pairing.

Obviously $\alpha_0=0$ for whatever $z$, so $\Delta_{AB,0}=0$, thus the electron paring is purely triplet. This demonstrates the spin-gauge locking effect of the Kitaev spin liquid, that a singlet in the spin space will be rendered by the PSG to a singlet in the SU(2) gauge space (seen from the expression of $\alpha_b$). However gauge charges can not condensed to a singlet state due to their bosonic nature, thus the single pairing is ruled out, as long as the quantum order persists.

The superconductivity transition temperature in the phase diagram shown with the dashed line is estimated as follows. At small doping, the phase stiffness $\rho_b=t_b\delta$ is proportional to doping, where $t_b=3t(u_0+u_a+2u_b)/8$, and $T_c$ is estimated from the Kosterlitz-Thouless transition\cite{KT} temperature $T_c=\pi\rho_b/2$. At large doping, the mean field gap is small which controls $T_c\sim \Delta_f$, where $\Delta_f\simeq J_\text{K}(u_0^2+u_a^2)^{1/2}(1-\delta^2)/4$. In between, we interpolate via the formula\cite{IoffeLarkin} $T_c^{-1}=(\pi\rho_b/2)^{-1}+\Delta_f^{-1}$. Note, due to the absence of a finite temperature transition of the two dimensional free bosons, a naive mean field transition temperature is not specified.

\subsection{Symmetry and Topological Properties}

The mean field Hamiltonian of the Kitaev spin liquid appears surprising at first, since the only Majorana fermion with extended hopping is $\chi^0$, the real part of $f_\uparrow$, which seems to single out one spin species and break the time reversal symmetry. Actually, this is a gauge artifact. The SU(2) rotations between $f_{i\sigma}$ and $c_{i\sigma}$ will restore the time reversal symmetry on the electron level for the spin liquid. However, the SU(2) gauge redundancy is parameterized by holon fields $b_{i\nu}$ and must be resolved as the holon condenses. So, as has been discussed from the PSG prospective, the holon condensation must break the time reversal symmetry spontaneously, leading to a class D superconductor\cite{ReadGreen}, denoted as SC$_1$, with uniform magnetization $\langle\bm{S}\rangle \sim z^\dagger \bm{\sigma} z$.

\begin{figure}[htbp]
\begin{center}
\subfigure[]{\includegraphics[height=0.11\textheight]{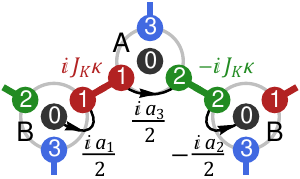}\label{fig:path}}\qquad
\subfigure[]{\includegraphics[height=0.11\textheight]{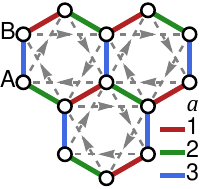}\label{fig:nnn}}
\caption{(Color online.) (a) The path of 3rd order perturbation. The 4 Majorana fermions on each site are denoted by their flavor indices. The effective second nearest neighbor hopping of $\chi^0$ fermion is bridged by two nearest-bond hopping of $\chi^{1,2,3}$. The on-site flavor changing process is assisted by the time-reversal-broken gauge potential. (b) Gray dashed arrows indicate the directions of  the second nearest neighboring bond.}
\end{center}
\end{figure}

Let us elaborate on the microscopic mechanism which gaps the $\chi^0$ Majoranas in SC$_1$. If we view the charge and spin as separate excitations, one may expect the same spectrum as the Kitaev spin liquid, i.e. gapless $\chi^0$ Majorana modes, to persist into the superconductor. However, the time reversal symmetry, which protects this gaplessness in the spin liquid, is lost in the superconductor. This can lead to an energy gap for $\chi^0$ (as shown in Fig.\,\ref{fig:disp1}), tied to the strength of the condensate. Because the uniform SU(2) gauge charge provided by the holon condensate offsets the SU(2) gauge potential $a^l$ ($l=1,2,3$) from zero, in order to preserve the overall gauge singlet condition. It is found that $a^l\simeq\delta J_\text{K}$ increases with doping. For small doping $\delta$, we treat $a^l/J_\text{K}$ as a perturbation. Integrating out the gapped Majorana modes $\chi^{1,2,3}$ generates next nearest neighboring (nnn) (Fig.\,\ref{fig:nnn}) hopping of $\chi^0$ fermions through a 3rd order perturbation correction, as illustrated in Fig.\,\ref{fig:path}. The effective Hamiltonian for $\chi^0$ reads
\begin{equation}\label{eq:Heff}
H_\text{eff}=J_K\eta\sum_{\langle i j\rangle}i \chi_i^0\chi_j^0+v\sum_{\langle\langle i j\rangle \rangle}i \chi_i^0\chi_j^0,
\end{equation}
where $v=a^1 a^2 a^3/(8J_\text{K}^2u_0^2)$ and $\langle\langle i j\rangle \rangle$ denotes the oriented nnn bond, with the bond direction specified in Fig.\,\ref{fig:nnn}.  According to the Kitaev spin liquid PSG (see Tab.\,\ref{tab:PSG}), the nnn coupling term is time reversal odd (since $i\to-i$), and is allowed only because time reversal symmetry is broken by the gauge charge condensation here.

The resulting $\chi^0$ Hamiltonian Eq.\,(\ref{eq:Heff}) is a Majorana version of the Haldane model\cite{HaldaneHoneycomb}. It is known that the nnn coupling gaps the Dirac cones and leaves one unit of Chern number in the ground state. This requires all $a^l$ to be nonvanishing. It is actually energetically favorable for the holon condensate (i.e. magnetization) to be in the (111) (or equivalent $(\pm 1 \pm 1 \pm 1)$) direction, corresponding to $a^1=a^2=a^3$ which maximizes the spinon gap $m=3\sqrt{3}|v|\sim|a^1 a^2 a^3|$. Therefore in the small doping limit, the ground state is a fully-gapped topological superconductor with $+1$ Chern number, which implies a gapless chiral Majorana edge mode and a Majorana zero mode in the vortex core. This is the same topology as a $p_x+ip_y$ superconductor of spin polarized fermions\cite{ReadGreen}; here the ``spin-polarization'' arises from the peculiar dispersion of fermions in the Kitaev spin liquid. At larger doping the Chern number changes, as shown in Fig.\,\ref{fig:phase}. The transition $+1\rightarrow-2$ in the SC$_1$ phase corresponds to a band gap closing at $M$ point due to the softening of $\chi^{1,2,3}$ modes.

\section{Discussion and Conclusion}

\subsection{Overdoped Regime and Weak Coupling BCS}

In the overdoped FL phase where correlations are weak, the superconductivity (SC$_2$) can be studied under the BCS paradigm by treating $H_\text{HK}$ as an interaction and decomposing it into the Cooper channel. In the small $J$ limit, The instability is found in the spin-triplet pairing channel, because the spin model is ferromagnetic. To first order in weak coupling both the time reversal invariant superconductor (the two dimensional analog of He$_3$ B phase) and the time reversal symmetry broken triplet superconductor (the analog of the He$_3$ A phase) are degenerate. To next order, the calculation in Ref.\,\onlinecite{BCS} showed that the time-reversal-invariant $p$-wave superconductor is preferred. Beyond weak coupling it is hard to decide which of these two possibilities is realized, a problem that is well known from He$_3$ physics\cite{leggett}. Here, our choice of PSG selects the time reversal ($T$) broken state, while a different choice would yield  the $T$ symmetric state. Therefore we mention both these possibilities as potentially relevant to the material at hand at high doping. In either case, the SC$_2$ phase is dominated by the Fermi liquid physics and is separated  by a first order transition from the spin-liquid-controlled time-reversal-broken SC$_1$ phase elaborated in this work. Because of the distinct underlying mechanism, its is not surprising that SC$_1$ and SC$_2$ can be quite different in many aspects.

\subsection{Conclusion}

A time-reversal-broken spin-triplet topological superconductor was found in the doped Kitaev spin liquid within the SU(2) slave boson formalism. 
A first order quantum transition around $\delta_c \sim J_\text{K}/t$ separates the spin triplet superconductor into two distinct classes: SC$_1$ (controlled by $J_\text{K}$) is governed by the spin liquid physics and reflects the underlying quantum order, while SC$_2$ (controlled by $t$) is a more conventional BCS-type superconductor. Although both ultimately trace their origins to the magnetic couplings, the detailed mechanisms are rather different.  This is in sharp contrast to the $t$-$J$ model in the context of cuprates, where, at least qualitatively, $d$-wave superconductivity is realized throughout.

A promising candidate material is A$_2$IrO$_3$ (A\,=\,Na, Li) \cite{Na2IrO3,Li2IrO3,Hill}, although experiments suggest magnetic ground state, rather than spin liquid. However, it has been argued that doped charges are more mobile in spin liquids, as compared to antiferromagnetic states where they interfere with the ordered pattern\cite{RVB}. Therefore one may hope that the results derived here also hold for magnetic ground states that are proximate to the Kitaev phase. Our main prediction is that doping these systems should lead to spin triplet topological superconductors with superconducting $T_c$ a fraction of the magnetic exchange. Assuming $J_\text{K}\sim$ 100\,-\,150K \cite{Na2IrO3,Li2IrO3,kimchiyou} a crude estimate of maximum superconducting transition temperature is 15\,-\,20K. Although we are not aware of doping studies on this class of materials, the related iridium perovskite Sr$_2$IrO$_4$, a 5d cuprate analog \cite{JK,Fa} has been doped in the bulk\cite{Cao}, and recent years have witnessed significant progress in doping techniques. We hope our results will spur future experiments in this direction.

\begin{acknowledgments}
We thank Hong Yao and Tarun Grover for helpful discussions and the Laboratory
Directed Research and Development Program of Lawrence Berkeley National Laboratory under US Department of Energy
Contract No. DE-AC02-05CH11231 for funding.  YZY was supported by the China Scholarship Council and Tsinghua Education Foundation in North America. IK is supported by the National Science Foundation Graduate Research Fellowship under Grant No. DGE 1106400.
\end{acknowledgments}

\appendix
\section{$\mathbb{Z}_2$ Projective Symmetry Group on Honeycomb Lattice}\label{app:PSG}

Here we present the classification of $\mathbb{Z}_2$ projective symmetry group (PSG) on Honeycomb lattice without spin rotational symmetry (but preserving time several symmetry). 144 solutions of algebraic PSG were found.

On the Honeycomb lattice, each unit cell is labeled by its integer coordinates $x_1$ and $x_2$ along the translation axes of $T_1$ and $T_2$. A spin site is further specified by its sublattice label $A$ or $B$ within the unit cell, see Fig.\,\ref{fig:lattice}. The symmetry group operators act on the lattice by
\begin{equation}\label{eq:PSG g}
\begin{split}
T_1(x_1,x_2) & = (x_1+1,x_2),\\
T_2(x_1,x_2) & = (x_1,x_2+1),\\
C_6(x_1,x_2,A) & = (x_1-x_2,x_1,B), \\ C_6(x_1,x_2,B) & = (x_1-x_2-1,x_1,A),\\
\sigma(x_1,x_2,A) & = (x_2,x_1,B), \\ \sigma(x_1,x_2,B) & = (x_2,x_1,A).
\end{split}
\end{equation}
The sublattice label is omitted if a formula holds in both sublattices. 
Later we will refer to the principal unit cell by omitting the unit cell index, i.e. $(0,0,A)\equiv(A)$, $(0,0,B)\equiv(B)$. The representation of symmetry operators in the spin space will be given after further discussion in Eq.\,(\ref{eq:PSG U}).

The symmetry group of a general spin model on the Honeycomb lattice is generated by 5 generators $\mathcal{T}$, $T_1$, $T_2$, $C_6$ and $\sigma$ with the following 13 definition relations
\begin{eqnarray}
&&T_1T_2T_1^{-1}T_2^{-1}=1,\label{eq:def T1T2}\\
&&\mathcal{T} T_1\mathcal{T} T_1^{-1}=\mathcal{T} T_2\mathcal{T} T_2^{-1}=1,\label{eq:def TT12}\\
&&C_6T_1C_6^{-1}T_1^{-1}T_2^{-1}=C_6T_2C_6^{-1}T_1=1,\label{eq:def CT12}\\
&&\sigma  T_1\sigma ^{-1} T_2^{-1}=\sigma  T_2\sigma ^{-1} T_1^{-1}=1,\label{eq:def sT12}\\
&&\mathcal{T}^2=C_6^6=\sigma ^2=1,\label{eq:def cyc}\\
&&\mathcal{T} C_6\mathcal{T} C_6^{-1}=1,\label{eq:def TC}\\
&&\mathcal{T} \sigma  \mathcal{T} \sigma ^{-1}=1,\label{eq:def Ts}\\
&&C_6\sigma  C_6\sigma =1.\label{eq:def Cs}
\end{eqnarray}
In general each definition relation takes the form of $\cdots g_2g_1=1$, where $\cdots g_2g_1$ denotes a sequence of symmetry group operations. Then according to Eq.\,(\ref{eq:PSG}), under the PSG operation, the spinon matrix $F_i$ transforms as
\begin{equation}
F_i\to U_{g_1}^{\dagger }U_{g_2}^{\dagger}\cdots F_{\cdots  g_2g_1(i)}\cdots G_{g_2}\left(g_1(i)\right)G_{g_1}(i).
\end{equation}
Because the bunch of operations $\cdots g_2g_1$ actually result in the identity operation, so they must not affect the spin degree of freedom: $U_{g_1}^{\dagger }U_{g_2}^{\dagger}\cdots =\sigma_0$ and must also restore the original lattice site: $\cdots  g_2g_1(i)=i$, hence the PSG operation becomes a pure gauge operation
\begin{equation}
F_i\overset{\cdots  g_2g_1}{\longrightarrow }F_i\cdots G_{g_2}\left(g_1(i)\right)G_{g_1}(i).
\end{equation}
All the pure gauge operations that leaves the mean field ansatz invariant constitute a subgroup of the PSG, known as the invariant gauge group (IGG). So we must have $\cdots G_{g_2}\left(g_1(i)\right)G_{g_1}(i)\in\text{IGG}$. Here we are interested in the classification of $\mathbb{Z}_2$ spin liquid, so we will focus on the case that $\text{IGG}=\mathbb{Z}_2$. Thus for each definition relation $\cdots g_2g_1=1$, there is a corresponding PSG representation
\begin{equation}\label{eq:PSG Rep}
\cdots G_{g_2}\left(g_1(i)\right)G_{g_1}(i) = \eta_m,
\end{equation}
where $\eta_m=\pm\sigma_0$ will be used to denote the sign factors hereon. For the 13 definition relations, we introduce 13 sign factors $\eta_1, \eta_2, \cdots, \eta_{13}$ to denote the corresponding IGG elements. In the following, we may write the PSG representation Eq.\,(\ref{eq:PSG Rep}) in short as $\cdots G_{g_2}G_{g_1}=\eta_m$ by omitting the site labels so as to save the space.

However special attention should be paid to the time reversal operation, because it involves the complex conjugate operator $\mathcal{K}$ which does not commute with $G_g$ in general. As can be seen from Eq.\,(\ref{eq:PSG TimeRev}), $\mathcal{K}$ must be placed right after each $G_\mathcal{T}$. For example, $\mathcal{T} C_6\mathcal{T} C_6^{-1}=1$ should be represented as
\begin{equation}
G_{\mathcal{T}}(i)\mathcal{K} G_{C_6}(i')G_{\mathcal{T}}(i')\mathcal{K} G_{C_6}^{-1}(i')=\eta _{11},
\end{equation}
where $i'=C_6^{-1}(i)$. Here we have used the rule that $G_{g^{-1}}(g(i))=G_g^{-1}(i)$ to simplify the inverse operations.

To classify the PSG's one should take care of the gauge redundancy in the solution of $G_g$. Two PSG's are gauge equivalent if their solutions of $G_g$ are related by a set of local SU(2) gauge transform $G_g(i)\to W_{g(i)}^\dagger G_g(i)W_i : W_i\in\text{SU(2)}$. To reduce the gauge redundancy, gauge fixing will be used while solving the equations of $G_g$. First of all, the relative gauge between the unit cells can be fixed by setting $G_{T_2}(x_1,x_2)=\sigma_0$, and $G_{T_1}(x_1,0)=\sigma_0$, then Eq.\,(\ref{eq:def T1T2}) can be represented as $G_{T_1}(x_1,x_2+1)=\eta _1G_{T_1}(x_1,x_2)$, which gives the solution for translations
\begin{equation}\label{eq:G T1 T2}
G_{T_1}(x_1,x_2)=\eta _1^{x_2},G_{T_2}(x_1,x_2)=\sigma _0.
\end{equation}

Substitute Eq.\,(\ref{eq:G T1 T2}) into the PSG representation of Eq.\,(\ref{eq:def TT12}): $G_{\mathcal{T}}\mathcal{K}G_{T_1}\mathcal{K}=\eta_2 G_{T_1}$ and $G_{\mathcal{T}}\mathcal{K} G_{T_2}G_{\mathcal{T}}\mathcal{K}=\eta _3G_{T_2}$, we obtain
\begin{equation}\label{eq:Rep T T1 T2}
\begin{split}
G_{\mathcal{T}}(x_1+1,x_2)\mathcal{K} G_{\mathcal{T}}(x_1,x_2)\mathcal{K}&=\eta _2,\\
G_{\mathcal{T}}(x_1,x_2+1)\mathcal{K} G_{\mathcal{T}}(x_1,x_2)\mathcal{K}&=\eta _3,
\end{split}
\end{equation}
while on the other hand, from $G_{\mathcal{T}}\mathcal{K}G_{\mathcal{T}}\mathcal{K}=\eta_8$, we know $\mathcal{K}G_{\mathcal{T}}(x_1,x_2)\mathcal{K}=\eta _8G_{\mathcal{T}}^{-1}(x_1,x_2)$, so Eq.\,(\ref{eq:Rep T T1 T2}) becomes
\begin{equation}
\begin{split}
G_{\mathcal{T}}(x_1+1,x_2)&=\eta _2\eta _8G_{\mathcal{T}}(x_1,x_2),\\
G_{\mathcal{T}}(x_1,x_2+1)&=\eta _3\eta _8G_{\mathcal{T}}(x_1,x_2).
\end{split}
\end{equation}
The solution is
\begin{equation}\label{eq:G T}
G_{\mathcal{T}}(x_1,x_2)=\eta _2^{x_1}\eta _3^{x_2}\eta _8^{x_1+x_2}G_{\mathcal{T}}(0,0).
\end{equation}
Similarly by inserting Eq.\,(\ref{eq:G T1 T2}) into the PSG representation of Eq.\,(\ref{eq:def CT12}): $G_{C_6}G_{T_1}=\eta _4G_{T_2}G_{T_1}G_{C_6}$, $G_{C_6}G_{T_2}=\eta _5G_{T_1}^{-1}G_{C_6}$, and Eq.\,(\ref{eq:def sT12}): $G_{\sigma }G_{T_1}=\eta _6G_{T_2}G_{\sigma }$, $G_{\sigma }G_{T_2}=\eta _7G_{T_1}G_{\sigma }$, we find
\begin{equation}
\begin{split}
G_{C_6}(x_1+1,x_2)&=\eta _1^{x_1-x_2}\eta _4G_{C_6}(x_1,x_2),\\
G_{C_6}(x_1,x_2+1)&=\eta _1^{-x_1}\eta _5G_{C_6}(x_1,x_2),\\
G_{\sigma }(x_1+1,x_2)&=\eta _1^{x_2}\eta _6G_{\sigma }(x_1,x_2),\\
G_{\sigma }(x_1,x_2+1)&=\eta _1^{x_1}\eta _7G_{\sigma }(x_1,x_2),
\end{split}
\end{equation}
whose solutions are
\begin{equation}\label{eq:G C s}
\begin{split}
G_{C_6}(x_1,x_2)&=\eta _1^{x_1(x_1-1)/2-x_1 x_2}\eta _4^{x_1}\eta _5^{x_2}G_{C_6}(0,0),\\
G_{\sigma }(x_1,x_2)&=\eta _1^{x_1 x_2}\eta _6^{x_1}\eta _7^{x_2}G_{\sigma }(0,0).
\end{split}
\end{equation}
However, it worth mention that $\eta_4$, $\eta_5$, $\eta_6$ and $\eta_7$ are not independent\cite{LuRan}. Because in their equations, either $G_{T_1}$ or $G_{T_2}$ only appears once, which means if we fix the inter-unit-cell gauge in a different way such that $G_{T_1}\to -G_{T_1}$ or  $G_{T_2}\to - G_{T_2}$, the above four $\eta$'s will be affected. But this does not affect the mean field ansatz, as all the ansatz are given in the bilinear form which are invariant under this $\mathbb{Z}_2$ gauge transform. Therefore, make use of this $\mathbb{Z}_2$ gauge freedom, one can set two out of the four $\eta$'s to identity, say $\eta_4 = \eta_5 = \sigma_0$.

Substitute Eq.\,(\ref{eq:G T1 T2}), Eq.\,(\ref{eq:G T}), and Eq.\,(\ref{eq:G C s}) into the PSG representation of the rest of the definition relations, we find some constrains between the $\eta$'s. For example, from Eq.\,(\ref{eq:def cyc}) one can obtain
\begin{equation}
G_{\sigma }(A)G_{\sigma }(B)=G_{\sigma }(B)G_{\sigma }(A)=\left(\eta _6\eta _7\right)^{x_1+x_2}\eta _{10}.
\end{equation}
The left-hand-side is independent of $(x,y)$, so must the right-hand-side be, therefore we must have $\eta_6\eta_7=\sigma_0$, which means $\eta_6 = \eta_7$. Similarly from Eq.\,(\ref{eq:def TC}) and Eq.\,(\ref{eq:def Ts}) we find $\eta_2=\eta_3=\eta_8$ and from Eq.\,(\ref{eq:def Cs}) we find $\eta_5=\eta_6$, so eventually $\eta_4=\eta_5=\eta_6=\eta_7=\sigma_0$.

Now all the $G_g(x_1,x_2)$ has been reduced to $G_g(0,0)$ with in a single unit cell, concluded as follows
\begin{equation}
\begin{split}
G_{T_1}(x_1,x_2)&=\eta _1^{x_2},\\
G_{T_2}(x_1,x_2)&=\sigma _0,\\
G_{\mathcal{T}}(x_1,x_2)&=G_{\mathcal{T}}(0,0),\\
G_{C_6}(x_1,x_2)&=\eta _1^{x_1(x_1-1)/2-x_1 x_2}G_{C_6}(0,0),\\
G_{\sigma }(x_1,x_2)&=\eta _1^{x_1 x_2}G_{\sigma }(0,0).
\end{split}
\end{equation}
The remaining task is to determine $G_{\mathcal{T}}(0,0)$, $G_{C_6}(0,0)$ and $G_{\sigma}(0,0)$ from the following equations
\begin{eqnarray}
&&G_{\mathcal{T}}(A)\mathcal{K}G_{\mathcal{T}}(A)\mathcal{K}=G_{\mathcal{T}}(B)\mathcal{K}G_{\mathcal{T}}(B)\mathcal{K}=\eta _8,\label{eq:eta8}\\
&&G_{\mathcal{T}}(B)\mathcal{K} G_{C_6}(A)G_{\mathcal{T}}(A)\mathcal{K} =\eta _{11}G_{C_6}(A),\label{eq:eta11A}\\
&&G_{\mathcal{T}}(A)\mathcal{K} G_{C_6}(B)G_{\mathcal{T}}(B)\mathcal{K} =\eta _{11}G_{C_6}(B),\label{eq:eta11B}\\
&&G_{\mathcal{T}}(B)\mathcal{K} G_{\sigma }(A)G_{\mathcal{T}}(A)\mathcal{K} =\eta _{12}G_{\sigma }(A),\label{eq:eta12A}\\
&&G_{\mathcal{T}}(A)\mathcal{K} G_{\sigma }(B)G_{\mathcal{T}}(B)\mathcal{K} =\eta _{12}G_{\sigma }(B),\label{eq:eta12B}\\
&&G_{\sigma }(A)G_{\sigma }(B)=G_{\sigma }(B)G_{\sigma }(A)=\eta _{10},\label{eq:eta10}\\
&&\left(G_{C_6}(B)G_{\sigma }(A)\right)^2=\left(G_{C_6}(A)G_{\sigma }(B)\right)^2=\eta _1\eta _{13},\label{eq:eta13}\\
&&\left(G_{C_6}(B)G_{C_6}(A)\right)^3=\left(G_{C_6}(A)G_{C_6}(B)\right)^3=\eta _1\eta _9.\hspace{28pt}\label{eq:eta9}
\end{eqnarray}
The solution of the above equations leads to 144 algebraic PSG's which will be classified below. There are only two remaining SU(2) gauge freedom: the local gauge transform on site $A$ or site $B$ in the unit cell.

We start from the solution of $G_{\mathcal{T}}$. Let $G_{\mathcal{T}}=a_0\sigma_0+i a_1\sigma_1+i a_2\sigma_2+i a_3\sigma_3$ be the most general form of a SU(2) matrix (with $a_\mu\in\mathbb{R}$). Plug into the left-hand-side of Eq.\,(\ref{eq:eta8}), one finds
\begin{equation}
\begin{split}
G_{\mathcal{T}}\mathcal{K}G_{\mathcal{T}}\mathcal{K}
=&(a_0^2+a_1^2-a_2^2+a_3^2)\sigma_0\\
&+2ia_2(a_3\sigma_1+a_0\sigma_2-a_1\sigma_3).
\end{split}
\end{equation}
So if $\eta_8=-\sigma_0$, the solution is $a_2=\pm1$, $a_0=a_1=a_3=0$, i.e. $G_{\mathcal{T}}=\pm i\sigma_2$. Note that $i\sigma_2\mathcal{K}$ as a whole is SU(2) gauge invariant, thus the remaining gauge freedoms are preserved (even though $\sigma_2$ seems to be a special direction). While if $\eta_8=\sigma_0$, the solution is $a_2=0$, $a_0^2+a_1^2+a_3^2=1$. One can choose $G_{\mathcal{T}}(A)=G_{\mathcal{T}}(B)=\sigma_0$. In this case, the SU(2) gauge freedoms on both sites are fixed.

\emph{Class (I)}: $\eta_8=-\sigma_0$. Then $G_{\mathcal{T}}(A)=i \sigma _2$, $G_{\mathcal{T}}(B)=i \eta _{14}\sigma _2$, where $\eta_{14}=\pm\sigma_0$ is a new sign factor. Substitute into Eq.\,(\ref{eq:eta11A},\ref{eq:eta11B},\ref{eq:eta12A},\ref{eq:eta12B}), one finds $\eta _{11}=\eta _{12}=-\eta _{14}$, with no restriction on $G_{C_6}$ and $G_\sigma$. Fix the relative gauge between sites $A$ and $B$ by $G_{\sigma }(A)=\sigma _0$, then from Eq.\,(\ref{eq:eta10}),  $G_{\sigma }(B)=\eta _{10}$. Plug into Eq.\,(\ref{eq:eta13}),
\begin{equation}\label{eq:47}
G_{C_6}(B)^2=G_{C_6}(A)^2=\eta _1\eta _{13}.
\end{equation}
According to the sign of $\eta_1\eta_{13}$, the class (I) is further divided into two subclasses.

\emph{Class (I)(A)}: $\eta _1\eta _{13}=\sigma _0$. Then the solution of Eq.\,(\ref{eq:47}) reads $G_{C_6}(A)=\sigma _0$, $G_{C_6}(B)=\eta _{15}$. Substitute into Eq.\,(\ref{eq:eta9}), one finds $\eta_{15}=\eta_{1}\eta_{9}$.

The solutions in the class (I)(A) are summarized as
\begin{equation}
\begin{split}
G_{\mathcal{T}}(A)&=i \sigma _2,\\
G_{\mathcal{T}}(B)&=i \eta _{14}\sigma _2,\\
G_{C_6}(A)&=\sigma _0,\\
G_{C_6}(B)&=\eta _1\eta _9,\\
G_{\sigma }(A)&=\sigma _0,\\
G_{\sigma }(B)&=\eta _{10},
\end{split}
\end{equation}
which are controlled by $\eta_1, \eta_9, \eta_{10}, \eta_{14}$, providing $2^4=16$ PSG's.

\emph{Class (I)(B)}: $\eta _1\eta _{13}=-\sigma _0$. Then from Eq.\,(\ref{eq:47}), the general solution of $G_{C_6}$ is a linear combination of $i\sigma_1,i\sigma_2,i\sigma_3$. Because the global gauge freedom has not been fixed, so using this freedom, one can set $G_{C_6}(A)=i\sigma_1$.  Further assume $G_{C_6}(B)=\eta_1\eta_9(i \sigma _1\cos  \theta _1+\left(i \sigma _2\cos  \theta _2+i \sigma _3\sin  \theta _2\right)\sin  \theta _1)$, and plug into Eq.\,(\ref{eq:eta9}), one finds $\cos  3\theta _1=1,\sin  3\theta _1=0$, whose solution is $\theta_1=0,\pm2\pi/3$, and there is no restriction on $\theta_2$.

The solutions in the class (I)(B) are summarized as
\begin{equation}
\begin{split}
G_{\mathcal{T}}(A)&=i \sigma _2,\\
G_{\mathcal{T}}(B)&=i \eta _{14}\sigma _2,\\
G_{\sigma }(A)&=\sigma _0,\\
G_{\sigma }(B)&=\eta _{10},\\
G_{C_6}(A)&=i \sigma _1,\\
G_{C_6}(B)&=\eta _1\eta _9 i \sigma _1\exp( i \sigma _2\theta _1e^{i \sigma _1\theta _2}),
\end{split}
\end{equation}
which are controlled by $\eta_1, \eta_9, \eta_{10}, \eta_{14}, \theta_1$, providing $2^4\times3=48$ PSG's. Here $\theta_2\in[0,2\pi)$ is a free angle.

\emph{Class (II)}: $\eta_8 = \sigma_0$. Then $G_{\mathcal{T}}(A)=G_{\mathcal{T}}(B)=\sigma_0$. Therefore Eq.\,(\ref{eq:eta11A},\ref{eq:eta11B},\ref{eq:eta12A},\ref{eq:eta12B}) become
\begin{equation}\label{eq:KK}
\begin{split}
\mathcal{K} G_{C_6}(A)\mathcal{K} &=\eta _{11}G_{C_6}(A),\\
\mathcal{K} G_{C_6}(B)\mathcal{K} &=\eta _{11}G_{C_6}(B),\\
\mathcal{K} G_{\sigma }(A)\mathcal{K} &=\eta _{12}G_{\sigma }(A),\\
\mathcal{K} G_{\sigma }(B)\mathcal{K} &=\eta _{12}G_{\sigma }(B).
\end{split}
\end{equation}
The general solution of $\mathcal{K}G_g\mathcal{K}=G_g$ is $G_g=e^{i\sigma_2\theta}$, while the general solution of $\mathcal{K}G_g\mathcal{K}=-G_g$ is $G_g=i\sigma_3e^{i\sigma_2\theta}$. According to the sign of $\eta_{11}$ and $\eta_{12}$, the class (II) is further divided into four subclasses.

\emph{Class (II)(A1)}: $\eta _{11}=\eta _{12}=\sigma _0$. Then the general solution of Eq.\,(\ref{eq:KK}) reads $G_{C_6}(A)=e^{ i \sigma _2\theta _1}$, $G_{C_6}(B)=\eta_1\eta_9 e^{ i \sigma _2\theta _2}$, $G_{\sigma }(A)=e^{ i \sigma _2\theta _3}$, $G_{\sigma }(B)=\eta_{10} e^{ i \sigma _2\theta _4}$. Then according to Eq.\,(\ref{eq:eta10}), $\theta_4=-\theta_3$. Substitute into Eq.\,(\ref{eq:eta13}), we obtain $e^{ 2i \sigma _2\left(\theta _2+\theta _3\right)}=e^{ 2i \sigma _2\left(\theta _1-\theta _3\right)}=\eta _1\eta _{13}$, which implies $e^{ 2i \sigma _2\left(\theta _1+\theta _2\right)}=\sigma _0$, then Eq.\,(\ref{eq:eta9}) can be reduced to $e^{i\sigma _2\left(\theta _1+\theta _2\right)}=\sigma _0$, thus $\theta_2 = -\theta_1$. While $\theta_1$ and $\theta_3$ are related by
\begin{equation}\label{eq:theta1theta3}
\theta _1=\theta _3+
 \left\{ 
\begin{array}{ll}
 0 & \eta _1\eta _{13}=\sigma _0, \\
 \pi /2 & \eta _1\eta _{13}=-\sigma _0.
\end{array}\right.
\end{equation}

The solutions in the class (II)(A1) are summarized as
\begin{equation}
\begin{split}
G_{\mathcal{T}}(A)&=\sigma _0,\\
G_{\mathcal{T}}(B)&=\sigma _0,\\
G_{C_6}(A)&=e^{i \sigma _2\theta _1},\\
G_{C_6}(B)&=\eta _1\eta _9e^{-i \sigma _2\theta _1},\\
G_{\sigma }(A)&=e^{i \sigma _2\theta _3},\\
G_{\sigma }(B)&=\eta _{10}e^{-i \sigma _2\theta _3},
\end{split}
\end{equation}
which are controlled by $\eta_1, \eta_9, \eta_{10}, \eta_{13}$, providing $2^4=16$ PSG's. Here $\theta_3$ can be any angle, and $\theta_1$ follows from Eq.\,(\ref{eq:theta1theta3}).
 
\emph{Class (II)(B1)}: $-\eta _{11}=\eta _{12}=\sigma _0$. The solution of $G_\sigma$ is the same as the class (II)(A1), however the general solution of $G_{C_6}$ becomes: $G_{C_6}(A)=i \sigma _3e^{ i \sigma _2\theta _1}$, $G_{C_6}(B)=-\eta_1\eta_9 i \sigma _3e^{ i \sigma _2\theta _2}$. From Eq.\,(\ref{eq:eta13}) one finds $\eta_1\eta_{13}=-\sigma_0$. While from Eq.\,(\ref{eq:eta9}), $e^{3 i \sigma _2\left(\theta _1-\theta _2\right)}=1$, thus $(\theta_1-\theta_2)=0,\pm2\pi/3$.

The solutions in the class (II)(B1) are summarized as
\begin{equation}
\begin{split}
G_{\mathcal{T}}(A)&=\sigma _0,\\
G_{\mathcal{T}}(B)&=\sigma _0,\\
G_{C_6}(A)&=i \sigma _3e^{ i \sigma _2\theta _1},\\
G_{C_6}(B)&=-\eta _1\eta _9i \sigma _3e^{ i \sigma _2\theta _2},\\
G_{\sigma }(A)&=e^{i \sigma _2\theta _3},\\
G_{\sigma }(B)&=\eta _{10}e^{-i \sigma _2\theta _3},
\end{split}
\end{equation}
which are controlled by $\eta_1, \eta_9, \eta_{10}, (\theta_1-\theta_2)$, providing $2^3\times3=24$ PSG's. Here $\theta_2$ and $\theta_3$ can be any angles, and $(\theta_1-\theta_2)=0,\pm2\pi/3$.

\emph{Class (II)(A2)}: $-\eta _{11}=-\eta _{12}=\sigma _0$. Then the general solution of Eq.\,(\ref{eq:KK}) reads $G_{C_6}(A)=i \sigma _3e^{ i \sigma _2\theta _1}$, $G_{C_6}(B)=-\eta_1\eta_9 i \sigma _3e^{ i \sigma _2\theta _2}$, $G_{\sigma }(A)=i \sigma _3e^{ i \sigma _2\theta _3}$, $G_{\sigma }(B)=-\eta_{10}i \sigma _3e^{ i \sigma _2\theta _4}$. Then according to Eq.\,(\ref{eq:eta10}), $\theta_3=\theta_4$. Substitute into Eq.\,(\ref{eq:eta13}), then combining with Eq.\,(\ref{eq:eta9}), one finds $\theta_2 = -\theta_1$, and $\theta_1$ and $\theta_3$ are related by Eq.\,(\ref{eq:theta1theta3}).

The solutions in the class (II)(A2) are summarized as
\begin{equation}
\begin{split}
G_{\mathcal{T}}(A)&=\sigma _0,\\
G_{\mathcal{T}}(B)&=\sigma _0,\\
G_{C_6}(A)&=i \sigma _3e^{ i \sigma _2\theta _1},\\
G_{C_6}(B)&=-\eta _1\eta _9i \sigma _3e^{ i \sigma _2\theta _1},\\
G_{\sigma }(A)&=i \sigma _3e^{ i \sigma _2\theta _3},\\
G_{\sigma }(B)&=-\eta _{10}i \sigma _3e^{ i \sigma _2\theta _3},
\end{split}
\end{equation}
which are controlled by $\eta_1, \eta_9, \eta_{10}, \eta_{13}$, providing $2^4=16$ PSG's. Here $\theta_3$ can be any angle, and $\theta_1$ follows from Eq.\,(\ref{eq:theta1theta3}).

\emph{Class (II)(B2)}: $\eta _{11}=-\eta _{12}=\sigma _0$. The solution of $G_{\sigma}$ is the same as the class (II)(A2), however the general solution of $G_{C_6}$ becomes $G_{C_6}(A)=e^{ i \sigma _2\theta _1}$, $G_{C_6}(B)=\eta_1\eta_9 e^{ i \sigma _2\theta _2}$. From Eq.\,(\ref{eq:eta13}), it is found that $\eta_1\eta_{13}=-\sigma_0$. And Eq.\,(\ref{eq:eta9}) gives $e^{3 i \sigma _2\left(\theta _1+\theta _2\right)}=1$, so $(\theta_1+\theta_2)=0,\pm2\pi/3$.

The solutions in the class (II)(B2) are summarized as
\begin{equation}
\begin{split}
G_{\mathcal{T}}(A)&=\sigma _0,\\
G_{\mathcal{T}}(B)&=\sigma _0,\\
G_{C_6}(A)&=e^{ i \sigma _2\theta _1},\\
G_{C_6}(B)&=\eta _1\eta _9e^{ i \sigma _2\theta _2},\\
G_{\sigma }(A)&=i \sigma _3e^{ i \sigma _2\theta _3},\\
G_{\sigma }(B)&=-\eta _{10}i \sigma _3e^{ i \sigma _2\theta _3},
\end{split}
\end{equation}
which are controlled by $\eta_1, \eta_9, \eta_{10}, (\theta_{1}+\theta_2)$, providing $2^3\times3=24$ PSG's. Here $\theta_2$ and $\theta_3$ can be any angles, and $(\theta_1+\theta_2)=0,\pm2\pi/3$.

Now all the 144 algebraic PSG's has been classified. Given Eq.\,(\ref{eq:PSG G}), one can check $\eta_8=G_{\mathcal{T}}\mathcal{K}G_{\mathcal{T}}\mathcal{K}=-\sigma_0$ and $\eta_1\eta_3=\left(G_{C_6}(A)G_{\sigma }(B)\right)^2=(-\sigma_{C_6}\sigma_\sigma)^2=-\sigma_0$, which match the criterion of the class (I)(B).  So the PSG of Kitaev spin liquid belongs to the class (I)(B) with $\eta_1=-\eta_9=-\eta_{10}=-\eta_{14}=\sigma_0$ and $\theta_1=2\pi/3$.

Finally our classification is related to the previous work\cite{LuRan} in the following table. The number of PSG's in the class (II)(ii)(B)($\beta$) was miscounted in Ref.\,\onlinecite{LuRan} as 24, which should be 8 instead.
\begin{table}[htdp]
\caption{Relation between the new classification and the previous one.}
\begin{center}
\begin{tabular}{c|c}
This work & Ref.\,\onlinecite{LuRan} \\
\hline
(I)(A) & (I)(A)\\
(I)(B) & (I)(B)\\
(II)(A1) & (II)(i)(A) + (II)(i)(B)($\alpha$)\\
(II)(B1) & (II)(ii)(B)($\alpha$)\\
(II)(A2) & (II)(ii)(A) + (II)(ii)(B)($\beta$)\\
(II)(B2) & (II)(i)(B)($\beta$)
\end{tabular}
\end{center}
\label{default}
\end{table}%

\section{Mean Field Decomposition}\label{app:MF}
Rewrite the hopping term on a single bond in terms of $F_i$ and $B_i$ matrices as
\begin{equation}
\begin{split}
\sum_{\sigma}\left(c_{i\sigma}^{\dagger}c_{j\sigma}+h.c.\right)
&=\text{Tr}\,\sigma_3 C_i^{\dagger }C_j\\
&=\frac{1}{2}\text{Tr}\,\sigma_3 B_i^{\dagger }F_i^{\dagger }F_jB_j.
\end{split}
\end{equation}
According to Eq.\,(\ref{eq:F=chi}),
\begin{equation}\label{eq:FF=chichi}
\begin{split}
&F_i^\dagger F_j=\frac{1}{2} \sum_{\alpha=0}^3 \chi_i^\alpha \chi_j^\alpha\sigma_0 +\\
&\quad\frac{1}{2}\sum_{\alpha=1}^3 \left(i \chi _i^0\chi _j^{\alpha }-i \chi _i^{\alpha }\chi _j^0+\sum_{\beta,\gamma=1}^3 i \epsilon ^{\alpha \beta \gamma }\chi _i^{\beta }\chi _j^{\gamma }\right)\sigma _{\alpha }.
\end{split}
\end{equation}
Here we may simplify the expression by dropping the second term, and use $F_i^\dagger F_j\simeq\sum_{\alpha=0}^3\chi_i^\alpha\chi_j^\alpha\sigma_0$. There are two reasons. First, consider the dihedral group $D_2=\{1, e^{i\pi S_i^1}, e^{i\pi S_i^2}, e^{i\pi S_i^3}\}$, which is a symmetry of the model Hamiltonian, and should not be broken in the spin liquid or Fermi liquid phase. Under these $D_2$ operations, Majorana fermions undergo sign changes, say for example $e^{i\pi S_i^3}$: $\chi_i^0\rightarrow \chi_i^0$, $\chi_i^1\rightarrow -\chi_i^1$, $\chi_i^2\rightarrow -\chi_i^2$, $\chi_i^3\rightarrow \chi_i^3$, which can be seen from the behavior of spin operators $S_i^1\rightarrow -S_i^1$, $S_i^2\rightarrow -S_i^2$, $S_i^3\rightarrow S_i^3$. Then any term that change the flavor of Majorana fermions acquires a minus sign under at least one of the $D_2$ operations. So the $D_2$ symmetry preserves the flavor of Majorana fermions, and terms like $\chi_i^0\chi_j^\alpha$ and $i \epsilon ^{\alpha \beta \gamma }\chi _i^{\beta }\chi _j^{\gamma }$ are not allowed. Secondly, in the time reversal broken phase like the superconducting phase, the $D_2$ symmetry is broken. But in this case the bosons condense to a state described by Eq.\,(\ref{eq:b=z}), which does not support any boson gauge current, i.e. $\text{Tr}\, \langle\sigma _3B_i^{\dagger }\sigma _\alpha B_j\rangle =0$ ($\alpha=1,2,3$). So the second term in Eq.\,(\ref{eq:FF=chichi}) can not make a contribution to the mean field Hamiltonian in any case, and thus can be neglected for the sake of simplicity. Therefore the electron hopping term can be written as
\begin{equation}\label{eq:tMF}
\begin{split}
H_t
&=-\frac{t}{4}\sum _{\langle ij\rangle }\sum_{\alpha=0}^3 i\chi_i^\alpha \chi_j^\alpha\text{Tr}\, \sigma _3B_i^{\dagger }\left(-i \sigma _0\right)B_j\\
&=-\frac{t}{4}\sum _{\langle i j\rangle }\sum _{\alpha =0}^3 i \chi _i^{\alpha }\chi _j^{\alpha }\sum _{\nu =1}^2 \left(i b_{i\nu }^{\dagger }b_{j\nu }+h.c.\right)
\end{split}
\end{equation}

For the Kitaev spin coupling term $H_J$, we first rewrite the spin operator to match Kitaev's convention by combining it with the neutral gauge charge $\bm{K}=0$,
\begin{equation}\label{eq:S=chichi}
S_i^{a }\rightarrow S_i^{a }+ K_i^{a }\simeq(i \chi _i^0 \chi _i^{a}) (1-\delta ).
\end{equation}
The single-occupancy projector $(1-\delta)$ is appended to project out the holon gauge charge terms in $K_i^{a }$. Physically $(1-\delta)$ represents the probability that one electron actually appears on site so that the spin operator can make a effect. Substitute Eq.\,(\ref{eq:S=chichi}) into $H_{J_K}$,
\begin{equation}\label{eq:JMF}
H_{J_K} = J_K(1-\delta )^2 \sum _{\langle ij\rangle}   i \chi _i^0\chi _j^0i \chi _i^a\chi _j^a,
\end{equation}
where $a$ denotes the type of the bond $\langle i j\rangle$.

Then by introducing the mean field parameters in Eq.\,(\ref{eq:uw}) and following the standard slave boson mean field approach, Eq.\,(\ref{eq:tMF}) and Eq.\,(\ref{eq:JMF}) can be decomposed to the mean field Hamiltonian Eq.\,(\ref{eq:HMF}) through Hubbard-Stratonovich transform, with additional Lagrangian multipliers to enforce the doping and SU(2) gauge constraints.

\section{Case of Finite $J$}\label{app:finiteJ}
Starting from the $t$-$J$-$J_\text{K}$ model, $H=H_t+H_{\text{HK}}$ with
\begin{equation}
\begin{split}
H_t&=-t\sum_{\langle ij\rangle\sigma}\mathcal{P}c_{i\sigma}^\dagger c_{j\sigma}\mathcal{P}+h.c.-\mu\sum_{i\sigma}c_{i\sigma}^\dagger c_{i\sigma},\\
H_{\text{HK}}&=J\sum _{\langle ij\rangle } \bm{S}_i\cdot \bm{S}_j-J_\text{K}\sum _{\langle ij\rangle } S_i^aS_j^a.
\end{split}
\end{equation}
Following the SU(2) slave boson theory, introducing the mean field parameters: $u_{ij}^\alpha= \langle i \chi_i^\alpha \chi_j^\alpha \rangle$, $w_{ij}^\nu= \langle i b_{i\nu}^\dagger b_{j\nu} \rangle$, one obtains the mean field Hamiltonian
\begin{equation}
\begin{split}
H_{\text{MF}}=&\sum_{\langle i j\rangle}\left(\sum_{\alpha=0}^3 U_{ij}^\alpha i \chi_i^\alpha \chi_j^\alpha + \sum_{\nu=1}^2 W_{ij}^\nu (i b_{i\nu}^\dagger b_{j\nu}+h.c.)\right)\\
&+\sum_i\left(\sum_{l=1}^3 a_i^l K_i^l-\mu\sum_{\nu=1}^2 b_{i\nu}^\dagger b_{i\nu}\right),
\end{split}
\end{equation}
where the fermion bond strength reads
\begin{equation}
\begin{split}
U_{ij}^\alpha=&-\frac{t}{4}\sum_{\nu=1}^2(w_{ij}^{\nu}+c.c.) \\
&+(1-\delta)^2 \Bigg(u_{ij}^0(J_\text{K}\delta_{\alpha a}-J_\text{H}(1-\delta_{\alpha 0}))\\
&+\sum_{\beta=1}^{3}u_{ij}^\beta(J_\text{K}\delta_{\beta a}-J_\text{H})\delta_{\alpha 0}\Bigg),
\end{split}
\end{equation}
and the boson bond strength reads
\begin{equation}
W_{ij}^\nu=-\frac{t}{4}\sum_{\alpha = 0}^3 u_{ij}^\alpha.
\end{equation}
The index $a$ denotes the type of bond $\langle ij \rangle$. The boson chemical potential $\mu$ is chosen such that $\sum_{i,\nu}\langle b_{i\nu}^{\dagger }b_{i\nu}\rangle =\delta N$. The gauge potential $a_i^l$ is adjusted to ensure the gauge singlet condition $\langle K_i^l\rangle = 0$.

The mean field phase diagram can be obtained by solving the mean field Hamiltonian $H_\text{MF}$ self-consistently. All phase diagrams contains SC$_1$ phase with Chern number $+1$ at small doping limit and SC$_2$ phase at large doping, separated by the first order transition at $\delta_c$. Tab.\,\ref{tab:phase} list the values of $\delta_c$ for different settings of $t$ and $J$. $J/J_\text{K}=1/8$ corresponds to $\alpha=0.8$ according to the convention $J=1-\alpha$, $J_\text{K}=2\alpha$. We conclude that small Heisenberg coupling will not affect the phase diagram much on the mean field level.
\begin{table}[htb]
\caption{Kitaev spin liquid-FL transition point. }
\label{tab:phase}
\begin{tabular}{c|c|l}
\hline
 $t/J_\text{K}$ & $J_\text{H}/J_\text{K}$ & $\delta _c$ \\
\hline
 & 0 & 0.064 \\
\cline{2-3}
 \raisebox{3pt}[0pt]{10} &1/8&  0.056 \\
\hline
 & 0 & 0.12 \\
\cline{2-3}
\raisebox{3pt}[0pt]{5} &1/8&  0.11 \\
\hline
 & 0 &  0.22 \\
\cline{2-3}
\raisebox{3pt}[0pt]{2} & 1/8 &  0.20\\
\hline
\end{tabular}
\end{table}

On the type-3 bond, parameterize the mean field ansatz by $u_{ij}^0= u_0$, $u_{ij}^1=u_{ij}^2=u_b$, $u_{ij}^3=u_a$, $w_{ij}^1+w_{ij}^2= w$. Then
\begin{equation}
U_{ij}^0=-\frac{tw}{2}+(1-\delta)^2((J_\text{K}-J)u_a-2Ju_b).
\end{equation}
The evolution of fermion mean field parameters with doping at zero temperature is shown in Fig.\,\ref{fig:upara}.
The first order transition between the Kitaev spin liquid and the Fermi liquid phases happens when $U_{ij}^0=0$ (at this point, the $\chi^0$ band becomes completely flat and can not gain more energy from the magnetic interaction). It is found that the mean field solution follows $w=\delta$ and $u_b\simeq t\delta/(3J_\text{K})$ at zero temperature, then the first order transition point $\delta_c$ can be roughly estimated from the equation
\begin{equation}
\frac{t\delta_c}{2}=(1-\delta_c)^2\left((J_\text{K}-J)u_a-\frac{2Jt\delta_c}{3J_\text{K}}\right).
\end{equation}
Considering the case of $J=0$ and large $t$, the transition point will be simply given by $\delta_c=2u_a J_\text{K}/t$, where the value of $u_a\sim 0.3$ can be determined by the mean field solution.

\end{document}